# Stimulated emission and absorption of photons in magnetic point contacts: toward metal-based spin-lasers

Yu. G. Naidyuk, O. P. Balkashin, V.V. Fisun, and I. K. Yanson
*B. I. Verkin Institute for Low Temperature Physics and Engineering National Academy of Sciences of Ukraine, 47 Lenin Avenue, 61103 Kharkiv, Ukraine*

A. M. Kadigrobov,[1,2] R. I. Shekhter,[1] and M. Jonson,[1,3,4]
[1]*Department of Physics, University of Gothenburg, SE-412 96 Göteborg, Sweden*
[2]*Theoretische Physik III, Ruhr-Universität Bochum, D-44801 Bochum, Germany*
[3]*SUPA, Department of Physics, Heriot-Watt University, Edinburgh EH14 4AS, Scotland, UK*
[4]*Division of Quantum Phases and Devices, School of Physics, Konkuk University, Seoul 143-701, Korea*

V. Neu, M. Seifert
*Leibniz-Institut für Festkörper- und Werkstoffforschung Dresden e.V., Postfach 270116, D-01171 Dresden, Germany*

S. Andersson and V. Korenivski
*Nanostructure Physics, Royal Institute of Technology, 10691, Stockholm, Sweden*

**Abstract**

Point contacts between high anisotropy ferromagnetic $SmCo_5$ and normal metal Cu are used to achieve a strong spin-population inversion in the contact core. Subjected to microwave irradiation in resonance with the Zeeman splitting in Cu, the inverted spin-population relaxes through stimulated spin-flip photon emission, detected as peaks in the point contact resistance. Resonant spin-flip photon absorption is detected as resistance minima, corresponding to sourcing the photon field energy into the electrical circuit. These results demonstrate fundamental mechanisms that are potentially useful for designing metallic spin-based lasers.

At the core of spin-electronics are magnetic nanostructures utilizing spin-polarized electron currents for achieving novel functionality for devices such as field sensors [1, 2] and magnetic random access memory elements [3, 4]. Of particular interest is the case where a highly non-equilibrium spin population can be created and used to produce radiation (emission of photons). Spin injection into semiconductors has been reported to result in emission of circularly polarized light [5, 6] and used to measure the spin-polarization of the injected current. In the case where the electron transition emitting a photon is between two levels that are spin-split by an external field [6], however, the semiconductor's energy gap plays no role – the spin splitting produces two sub-bands within the conduction band. Based on this observation, some of us have proposed a new concept for a solid-state THz laser based on a population inversion of spin-split levels in a ferromagnetic metal [7]. The spin population is inverted by injecting spin-polarized currents through a tunnel barrier or a nano-constriction in such a way that electrons are either pumped in to the upper spin-split level or pumped out of the lower one. When the injection region is resonantly irradiated, stimulated photon emission takes place and is predicted to reach giant optical gain levels due to the vastly higher electron density in metals compared to semiconductors. If the photon emission rate can be made to exceed the photon absorption rate, the metal becomes transparent to the radiation and acts as a laser. The role of the energy gap in semiconductor-based lasers is played by the exchange- or Zeeman spin-split energy levels in the metallic active region for ferromagnetic (F) and normal metal (N) based lasers, respectively. The realization of such spin-based metallic THz and microwave lasers and masers would be a breakthrough in the field. A recent experiment using this new laser concept [7] has been interpreted in terms of photon emission by spin-flip relaxation of exchange-split spin levels in a ferromagnetic metal [8]. The resonant character of the detected THz radiation remains to be demonstrated, however.

In this paper we focus on the case where the active region is a normal metal with electron spin levels Zeeman-split by an external magnetic field and the corresponding radiation frequency in the microwave range. Magnetic point contacts (PCs) provide the ideal system for demonstrating the proposed effects, making possible very high injection current densities (~$10^9$ A/cm$^2$) into very small active volumes (1-10 nm diameter). Furthermore, point contact spectroscopy [9] allows in-situ monitoring of the relaxation processes taking place in the contact core. Using a high anisotropy spin injector, SmCo$_5$, we significantly extend our earlier result [10] on *stimulated emission of microwave radiation* by a metallic PC and demonstrate the mirror effect of spin-flip *photon absorption*. In addition we extend the modeling of these spin-photo-electronic effects by incorporating into the theory of [10] realistic non-photon spin relaxation in the active region.

SmCo$_5$ epitaxial films were prepared by pulsed laser deposition on Cr-buffered MgO(110) substrates with the *c*-axis (magnetic easy axis) aligned along one in-plane substrate edge. Film preparation and crystallographic and magnetic properties are comparable to samples described in [11], and the Cr capping layer was replaced by a Cu capping layer of about 3 nm in thickness. The SmCo$_5$ films were 50 nm thick, with the rms roughness of approximately 3 nm. They possess extremely large uniaxial magnetocrystalline anisotropy ($K_u \approx 10$ MJ/m$^3$) and a saturation polarization of about 1 Tesla, comparable to the respective values in single crystals, and display a sharp magnetization switching along the easy axis at coercive fields of 3 to 6 Tesla. Thus, applying a reversing field of up to 1 Tesla leaves the magnetization unchanged, while producing a strong *inverse* Zeeman splitting in a normal-metal PC to the surface of the film. Such tip-surface type PCs were produced directly in liquid-helium using a precision positioning mechanism, making it possible to in-situ change the location of the surface contact and the clamping force acting on the tip. The resistance of the contacts was in the range 10-30 Ω, corresponding to a PC diameter of 10-5 nm [9]. The differential resistance of the contacts, dV/dI(V), was recorded using the conventional technique of measuring the amplitude of the first harmonic of a low- frequency (443 Hz) modulation current superposed on a DC voltage bias. Magnetic fields of –5 to 5 T in the film plane were produced using a superconducting solenoid. Negative polarity of the bias voltage corresponds to electron injection from the tip into the film. A microwave field was applied to the PCs using a 10x23 mm X-band waveguide with a smooth transition to a 2x23 mm cross section, as illustrated in Fig. 1. The change in the PC resistance caused by microwave irradiation, $V_{det}$, was measured using a lock-in technique where the microwave power was modulated at low frequency taken as the reference for $V_{det}$ [12].

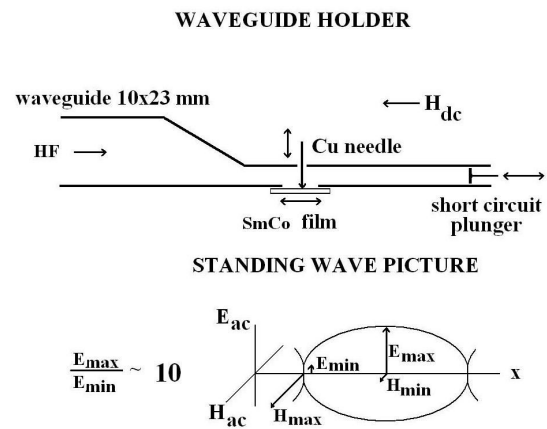

**Fig. 1.** Schematic of the waveguide PC holder, showing the position of the sample (film) and the tip (Cu). A short circuit plunger is used to tune the standing wave maxima to the PC position.

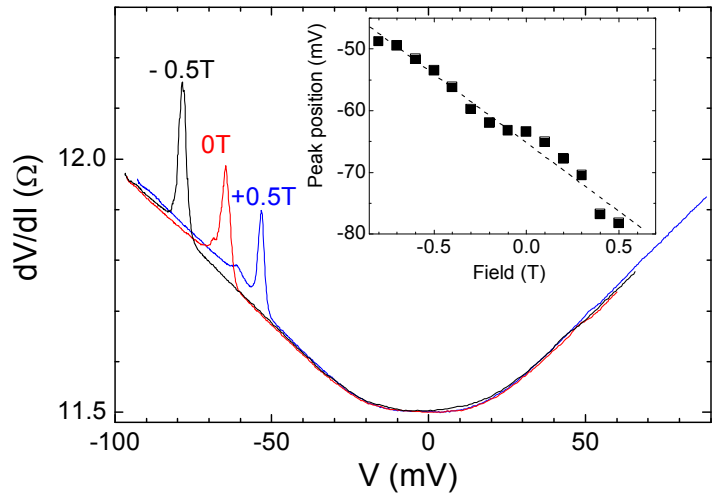

**Fig. 2** Differential resistance of a SmCo$_5$/Cu point contact with R= 11.5 Ω versus bias voltage at three magnetic fields. The inset shows the position of the STT peak vs magnetic field, with the dashed line representing a linear fit.

We first calibrate the obtained SmCo$_5$/Cu point contacts with respect to the now well understood effect of spin-transfer-torque (STT) driven magnetization dynamics caused by the secondary spin polarization in the PC core [12-14]. At high bias voltages applied to a PC, typically in access of 50 mV for 5-10 nm contacts (20 Ω contact resistance range), the spin accumulation at the F/N interface can become sufficiently high to cause a spin-wave instability (magnon excitation) and lead to a magnetization precession in the PC core, which is manifest through peaks in the differential resistance for the electron flow from the normal metal into the ferromagnet (negative bias voltage), such as shown in Fig. 2. The position of these peaks is dictated by the effective anisotropy acting on the ferromagnetic contact core, which in the case of SmCo$_5$ is the high intrinsic anisotropy of the material. A moderate applied magnetic field acts to enforce or compensate for this effective anisotropy, which results in shifting the position of the STT peak along the bias voltage direction – more bias and therefore a stronger spin-torque is required to excite the core magnetization subject to a higher anisotropy, and less spin-torque is needed when the anisotropy is reduced by the opposing field. Thus, the applied field of 0.5 T shifts the STT maximum by approximately a tenth along the voltage axis, which is to be expected for the intrinsic anisotropy of the material of several tesla. The STT peak position as a function of the applied field is shown in the inset to Fig. 2. The linear dependence observed agrees well with the expected behavior for STT dynamics and, importantly, the peak continues to be observed and its position changes monotonically as the field changes sign. These results show that the current through the SmCo$_5$/Cu point contacts under study is spin-polarized and produces a significant spin-accumulation in the contact core. Below we use this strong spin accumulation to demonstrate a new regime in magnetic PCs and a new effect based on photon-emitting and photon-absorbing spin-flip transitions, rather than magnon excitation responsible for the STT effect.

Figure 3 illustrates four basic spin-flip photon emission/absorption processes in a magnetic PC. For a majority ferromagnetic injector, such as SmCo$_5$, the electrons injected into the normal metal (N) have their magnetic moment along the magnetization M of the ferromagnet (F). With the external field applied anti-parallel to M, as in Fig. 3a, the electrons injected into the normal metal populate the high-energy Zeeman level there, which leads to an inverse population of the two spin-split levels. This inverse population is not in equilibrium but maintained by the bias current through the PC. Intrinsic spin relaxation in such weak spin-orbit nonmagnetic metals as Cu is relatively slow, so spin-flip relaxation through emission or absorption of photons can be dominant. Figure 3b shows the same photon emission process, but now for an electron flow from the normal metal to the ferromagnet and an external field applied parallel to the magnetization. The majority-spin subband in the ferromagnet contains more electron states than the minority-spin subband and hence majority-spin electrons carry most of the current through the N/F interface. This preferentially depopulates the lower-energy Zeeman level in the normal metal and thereby creates a spin population inversion, which can decay through photon emission. Photon absorption by a spin-flip process is illustrated in Fig. 3c, where the external field is anti-parallel to M and no bias is applied to the contact. The equilibrium, equally populated spin levels in the normal metal are subject to a microwave field with the frequency in resonance with the Zeeman splitting. Absorption of a microwave photon by an electron induces a spin-flip transition from the low-energy to the high-energy Zeeman level, which overpopulates the high-conductivity majority spin channel of the N/F interface and thereby produces a photo-current. A variation on the photon emission schemes of Figs. 3(a,b) is illustrated in Fig. 3d, where the field and magnetization are parallel and the injector is a minority ferromagnet, such the Fe-Cr alloy [15]. The injected electrons, now having magnetic moments opposite to M, populate the high-energy Zeeman level in the normal metal and the resulting spin population inversion can lead to photon emission [10]. It is clear that the majority injector schemes require a high anisotropy material, whereas the minority injector configuration can be implemented with a soft ferromagnet as well. Below we demonstrate these fundamental spin-photo-electronic effects using 5-10 nm scale PCs and a high anisotropy spin-majority injector.

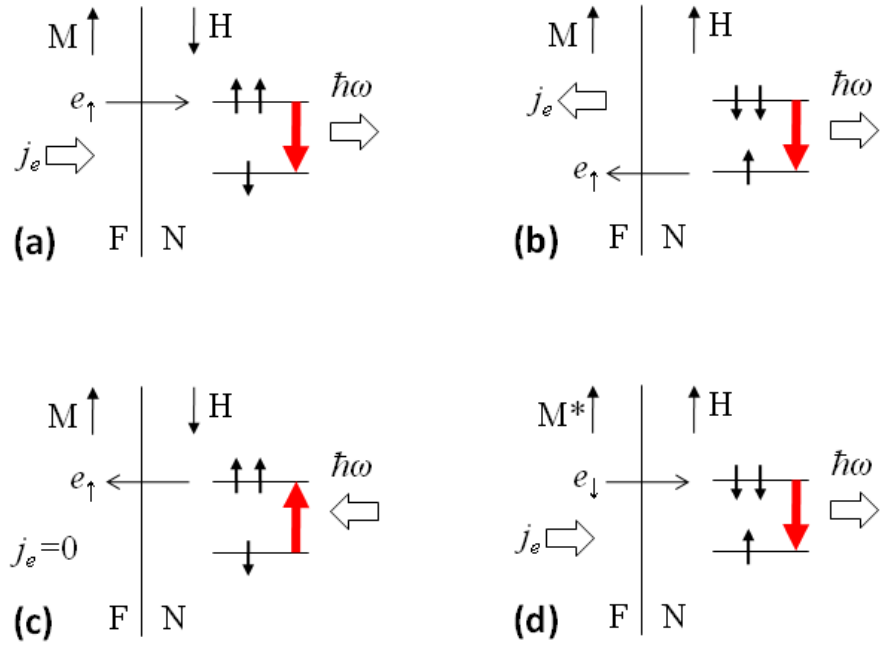

**Fig. 3** Schematic of four spin-flip transitions accompanied by photon emission or absorption: (a) magnetization M and applied field H are antiparallel, majority electrons in the ferromagnet F are injected into the normal metal N, leading to an inverse population of the Zeeman levels in N. A spontaneous or stimulated spin-flip transition emits a photon with frequency corresponding to the spin-splitting; (b) M and H are parallel, electrons are injected from N predominantly into the majority band of F resulting in inversely populated Zeeman levels in N; (c) M and H are antiparallel, no voltage bias is applied to the contact, absorption of photons at the frequency of the spin-splitting in N produces an overpopulation of electrons having moments along the high-conductivity majority spin-channel of F and, thereby, a flow of electrons through the interface (a photocurrent); (d) F is of the minority type as indicated by the asterisk, M and H are parallel, a spin-minority current inversely populates the Zeeman levels in N, which undergo spin-flip relaxation through photon emission. Small arrows denote the direction of the electron's magnetic moment.

We start by demonstrating the photon absorption effect illustrated in Fig. 3c. This configuration requires that a magnetic field, of sufficient strength to achieve a Zeeman splitting of reasonable magnitude, is anti-parallel to the magnetization of the injector. Therefore, a very high anisotropy material, such $SmCo_5$ with the coercive field of over 3 T, is required so that the magnetization would remain unchanged in the opposing magnetic field. Fig. 4 shows the voltage detected in synch with the modulated microwave irradiation ($V_{det}$, see experimental section above) acting on $SmCo_5$/Cu point contacts. The data are shown for two different PCs, with resistances of 24 and 29 Ω, corresponding to a contact diameter of approximately 5 nm. The microwave power is focused at the contact and is estimated to be of sufficient magnitude (photon density) to effectively induce spin-flip electron transitions in the contact core from the low- to the high-energy Zeeman level. This overpopulates the high-energy level, thereby producing a photo-voltage (and a photo-current in the measuring circuit). The core, being a few nanometers small, dominates the PC resistance so even small changes in the contact core resistance can be sensitively detected. The data shows a pronounced minimum when the field and the corresponding spin-splitting in the normal metal matches the photon energy (approximately 0.5 T for 14 GHz). These pronounced minima are absent in the absence of the microwave irradiation. The minima are unipolar in field (the signal is at the noise floor at H = -0.5 T), which is expected for the spin-flip process of Fig. 3c and leads to the conclusion that the minima are due to photon absorption. The minima in Fig. 4 are measured without a bias voltage applied to the contact, i.e., no spin injection driven by the circuit. The fact that the detector voltage is lower at the resonance, corresponding to a lower resistance at the resonance, means that the PC effectively acts as a source drawing power from the microwave field. Based on these observations we conclude that the measured

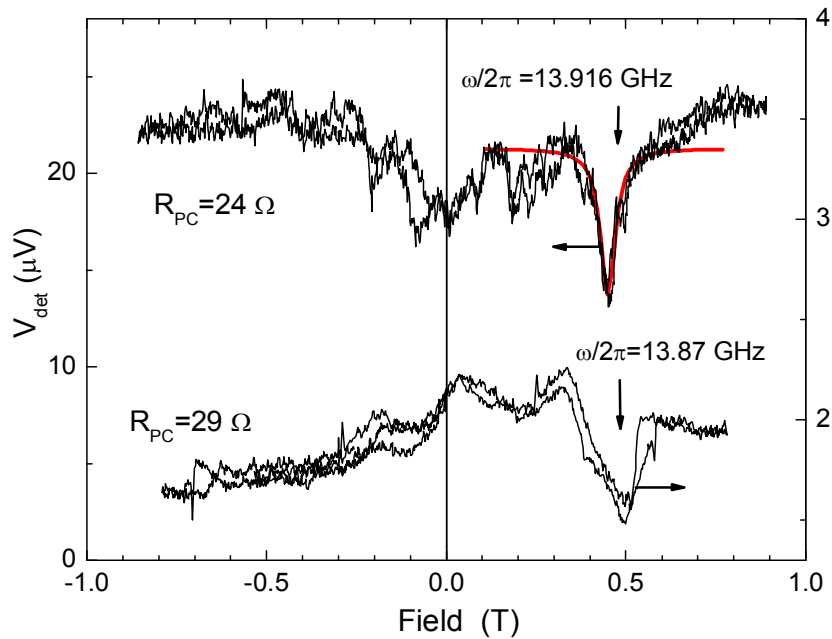

**Fig.4.** Detected voltage $V_{det}$ (see text) versus magnetic field for two unbiased $SmCo_5$/Cu point contacts subject to 13.9 GHz microwave irradiation at T=4.2 K. Vertical arrows indicate resonant field values $H_{res}$ for which the photon energy $\hbar\omega=2g\mu_s H_{res}$, if g=2 and s=1/2. Minima in $V_{det}$ correspond to minima in the PC resistance. The red (smooth) curve is a theoretical fit according to Eq. (1).

Zeeman-resonance minima are due to photon absorption by Zeeman-split spin-flip transition in the studied F/N point contacts.

We next demonstrate stimulated photon emission in SmCo$_5$/Cu point contacts in the configuration of Fig. 3b, where the external field is directed along the magnetization and the bias applied injects electrons from the normal metal into the ferromagnet. The amplitude of the bias voltage is kept relatively low (-20 mV), much lower than the -100 mV range needed to excite the STT magnon instabilities (see Fig. 2). The data for V$_{det}$ shown in Fig. 5 have a pronounced peak at approximately -0.33 T. The negative field is parallel to the magnetization, as required by the diagram of Fig.3b, and the peak is unipolar, which supports the interpretation that its origin is the spin-flip mechanism in question. The peak is absent in the absence of the microwave irradiation, which indicates that it is stimulated in nature.

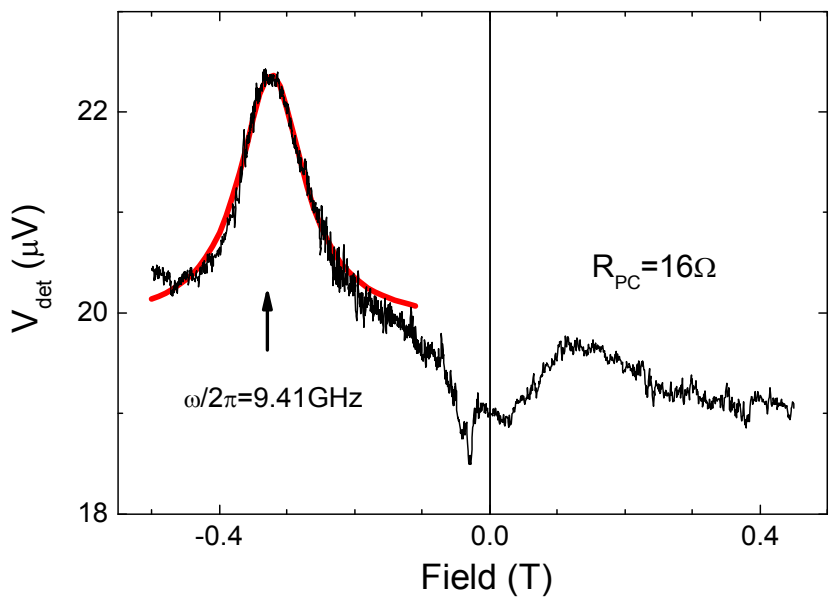

**Fig.5.** Detected voltage V$_{det}$ (see text) vs. magnetic field (average of two field scans) for an SmCo$_5$/Cu point contact under microwave irradiation at 4.2 K and biased with -20 mV (electrons injected into F). The arrow indicates the resonant magnetic field H$_{res}$ for which the photon energy $\hbar\omega = 2g\mu_s H_{res}$, if g=2 and s=1/2. The maximum of V$_{det}$ at the resonance corresponds to a maximum in the PC resistance. The red (smooth) curve is a theoretical fit according to Eq. (1).

We point out that the STT excitations observed at much high bias conditions do not require irradiation and can be observed for both polarities of the field, which rules them out as an explanation of the effect observed in Fig. 5. Furthermore, the fact that V$_{det}$ and therefore R$_{PC}$ have a maximum indicates that the effective resistance of the PC core increases at resonance, which is the expected behavior for a source dissipating power through photon emission when the Zeeman resonance condition is fulfilled. Based on these experimental observations we concluded that the effect responsible for the behavior in Fig. 5 is stimulated photon emission by spin-flip relaxation of inversely populated Zeeman levels in a normal metal, spin-pumped by a polarized current from a ferromagnetic injector.

In order to quantitatively model the photon resonance peaks observed, we extend our recent theory [10] to take into account spin relaxation in the PC core, which is of non-photonic origin. To do this we take the energy δ-function in the electron-photon collision integral (which determines the energy conservation law, see Eq. (2) in Ref. [10]) to be broadened due to spin-flip processes in the normal metal and find the relative change of the PC resistance:

$$\frac{\Delta R}{R} = \frac{4\pi\beta_{tr}^2}{3}\frac{c}{v_F}\frac{(\mu_B H_{max})^2}{\varepsilon_F \hbar v_{sf}}\left(n_0 \Omega_{pc}\right)\left(\frac{2e^2}{h}R\right)\arctan\frac{2\xi}{1-\xi^2+((\hbar\omega-2\mu_B H)/\hbar v_{sf})^2} \quad (1)$$

Here $\beta_{tr}$ is the transport parameter [10], the experimental value of which is ≈ 0.3, $c$ and $v_F$ are the light and the electron Fermi velocities, respectively, $\mu_B$ is the Bohr magneton, $H_{max}$ is the magnetic component of the irradiating electromagnetic field, $\varepsilon_F$ is the normal metal Fermi energy, $n_0$ is the electron density, $\Omega_{pc}$ is the PC volume, $\xi = \hbar\omega v_F / c v_{sf}$, and $v_{sf}$ is the spin-flip rate due to spin-orbit or impurity scattering. Using $v_{sf}$ as the fitting parameter we have fitted the above equation to the experimental data and found good agreement for $v_{sf} = 2 \cdot 10^9$ s$^{-1}$ for the photon absorption peak of Fig. 4 (top curve) and $v_{sf} = 1 \cdot 10^{10}$ s$^{-1}$ for the stimulated photon emission peak of Fig. 5. These values correspond to a spin relaxation time of 100-500 picoseconds, which agrees well with the typical spin relaxation time in our similarly prepared normal metal films and F/N nanodevices, measured using non-local spin injection techniques [16, 17], as well as with data in the literature [18]. The fact that the photon peaks are broader in the case of spin pumping and emission compared to absorption can be due to the non-equilibrium situation in the spin-pumped case, where the injected hot electrons have not fully equilibrated in energy into the respective Zeeman levels before undergoing spin-flip. The absorption, on the other hand, takes place from the equilibrium Zeeman-split distributions at small or zero bias.

A relatively small number of PCs, ~1% of the total number of tested contacts, showed the spectroscopic features of photon emission and absorption. This is to be compared to approximately 5-10% of PCs with STT features. The relatively low observation rate is potentially related to more stringent requirement on the spin accumulation in the PC contact core in the case of photonic transition, compared to the STT case. An imperfect F/N interface can lead to a significant spin-flip scattering exciting magnons at the surface of the ferromagnet (STT) while at the same time suppressing the spin accumulation in normal metal contact core and thereby spin-flip photonic transition. Further progress toward spin-laser effects in metallic magnetic nanostructures should come from optimizing the microstructure and geometry of the samples and improving the efficiency of the spin-injector electrodes.

In conclusion, strong spin accumulation in nanometer sized PCs is used to achieve spin population inversion and demonstrate the mechanisms of stimulated spin-flip photon emission and photon absorption. These effects can form the foundation for spin-lasers, with the active region being a metal.

**Acknowledgements**

Financial support from the European Commission (FP7-ICT-FET project no. 225955 STELE), the Swedish VR, and the Korean WCU program funded by MEST/NFR (R31-2008-000-10057-0) is gratefully acknowledged. We thank V. Pashchenko for measuring the magnetic properties of the film samples.

**References**

[1] N. Baibich , J. M. Broto, A. Fert, F. Nguyen Van Dau, F. Petroff, P. Eitenne, G. Creuzet, A. Friederich, and J. Chazelas, Phys. Rev. Lett. **61**, 2472 (1988).

[2] B. Dieny, V. S. Speriosu, S. S. P. Parkin, B. A. Gurney, D. R. Wilhoit, and D. Mauri, Phys. Rev. B **43**, 1297 (1991).

[3] J. S. Moodera, Lisa R. Kinder, Terrilyn M. Wong, and R. Meservey, Phys. Rev. Lett. **74**, 3273 (1995).

[4] W. J. Gallagher, S. S. P. Parkin, Y. Lu, X. P. Bian, A. Marley, K. P. Roche, R. A. Altman, S. A. Rishton, C. Jahnes, T. M. Shaw, and G. Xiao, J. Appl. Phys. **81**, 3741 (1997); B. N. Engel, J. Akerman, B. Butcher, R. W. Dave, M. DeHerrera, M. Durlam, G. Grynkewich, J. Janesky, S. V. Pietambaram, N. D. Rizzo, J. M. Slaughter, K. Smith, J. J. Sun, and S. Tehrani, IEEE Trans. Magn. **41**, 132 (2005).

[5] H. B. Zhao, D. Talbayev, G. Lüpke, A. T. Hanbicki, C. H. Li, M. J. van't Erve, G. Kioseoglou, B. T. Jonker, Phys. Rev. Lett. **95**, 137202 (2005), and references therein.

[6] N. A. Viglin, V. V. Ustinov, and V. V. Osipov, JETP Lett. **86**, 193 (2007).

[7] A. Kadigrobov, Z. Ivanov, T. Claeson, R. I. Shekhter, and M. Jonson, Europhys. Lett., **67**, 948 (2004).

[8] Yu. V. Gulayev, P. E. Zilberman, S. G. Chigarev and E. M. Epshtein, J. Commun. Technol. Electron. **55**, 1132 (2010). (arXiv:1004.5248).

[9] Y. G. Naidyuk and I. K. Yanson, Point-Contact Spectroscopy (Springer Series in Solid-State Sciences, vol. 145) (New York: Springer) (2005).

[10] A. M. Kadigrobov, R. I. Shekhter, S. I. Kulinich, M. Jonson, O. P. Balkashin, V. V. Fisun, Yu. G. Naidyuk, I. K. Yanson, S. Andersson, V. Korenivski, New J. Phys. **13,** 023007 (2011).

[11] A. Singh, V. Neu, S. Fähler, K. Nenkov, L. Schultz, and B. Holzapfe, Phys. Rev. B **77**, 104443, (2008).

[12] O. P. Balkashin, V. V. Fisun, I. K. Yanson, L. Y. Triputen, A. Konovalenko, and V. Korenivski, Phys. Rev. B **79,** 092419 (2009).

[13] Y. Ji, C. L. Chen, and M. D. Stiles, Phys. Rev. Lett. **90**, 106601 (2003).
T. Y. Chen, Y. Ji, C. L. Chen, and M. D. Stiles, J. Appl. Phys. **97**, 10C709 (2005).

[14] I. K. Yanson, Yu. G. Naidyuk, D. L. Bashlakov, V. V. Fisun, O. P. Balkashin, A. Konovalenko, V. Korenivski, R. I. Shekhter, Phys. Rev. Lett. **95**, 186602 (2005);
I. K. Yanson, Yu. G. Naidyuk, V. V. Fisun, A. Konovalenko, O. P. Balkashin, L. Yu. Triputen, and V. Korenivski, Nano Lett. **7**, 927 (2007).

[15] C. Vouille, A. Barthélémy, F. Elokan Mpondo, A. Fert, P. A. Schroeder, S. Y. Hsu, A. Reilly, and R. Loloee, Phys. Rev. B **60**, 6710 (1999).

[16] N. Poli, J. P. Morten, M. Urech, A. Brataas, D. B. Haviland, V. Korenivski, Phys. Rev. Lett. **100**, 136601 (2008).

[17] M. Urech, V. Korenivski, N. Poli, and D. B. Haviland, Nano Lett. **6**, 871 (2006).

[18] J. Bass and W. P. Pratt Jr., J. Phys. Cond. Matt. **19**, 183201 (2007).